\documentstyle[12pt]{article}
\begin{document}
\begin{center}
SYNCHRONIZATION REGIMES IN SYSTEMS WITH MULTIPLE TIME-DELAYS\\
E.M.Shahverdiev \footnote {e-mails:eshahver@ictp.trieste.it;shahverdiev@physics.ab.az}\\
Institute of Physics, 33,H.Javid Avenue, Baku-370143, Azerbaijan\\
Abdus Salam ICTP, Strada Costiera 11-34014, Trieste, Italy\\
~\\
ABSTRACT\\
\end{center}
We present the first report on chaos synchronization between two unidirectionally 
linearly and nonlinearly coupled systems with multiple time-delays and using 
the Razumikhin-Lyapunov approach find the existence and stability conditions for 
different synchronization regimes. The approach is tested on the famous nonlinear 
models-Ikeda, Mackey-Glass and  Lang-Kobayashi systems.\\
PACS number(s):05.45.Xt, 05.45.Vx, 42.55.Px, 42.65.Sf\\
\indent {\it Introduction.}- There is continued growth in the field of chaos control [1] 
and ever-increasing appreciation of its appications among researchers. Application of 
chaos control can be found in secure communication, optimization of nonlinear system 
performance, modeling brain activity and pattern recognition phenomena [2], 
species population control [3],etc.\\
\indent Recently delay differential equations (DDEs) have attracted a lot of 
attention in the field of nonlinear dynamics. DDEs are used to model dynamical systems 
in many scientific and enginering areas,e.g.optics, biology, climatology, economy, 
cryptosystems based on synchronized hyperchaos, networks [4-5],
etc. In comparison with a single time-delay DDEs with multiple time-delays are 
more relaistic models in the interacting complex systems. These delays result naturally 
from the finite propagation velocity of information, from the latency of feedback loops, 
from the finite switching times between different states of the system.
Additional time-delays could be useful e.g. to stabilize 
nonlinear system's output [6].The application possibilities based on chaos require 
proper control of complexity. To the best our knowledge chaos synchronization  
between the systems with multiple time-delays has not been investigated yet. 
Having in mind enormous application implications of chaos synchronization e.g. in secure 
communication, performance optimization in nonlinear systems, stabilization problems, 
etc. investigation of synchronization regimes in the multiple time-delay systems is of 
paramount importance.\\
\indent In this Letter we present the first report on chaos synchronization between 
two unidirectionally linearly and nonlinearly coupled chaotic systems with mutiple 
time-delays and find the existence and stability conditions for different synchronization 
regimes. We test the approach on the paradigm Ikeda, Mackey-Glass and Lang-Kobayashi models [4]. 
We hope this research will pave the way for the intensive experimental investigations 
of chaos synchronization in the multple time-delays systems.\\
\indent {\it General approach.}-Consider synchronization between the double-feedback systems of general form ,
$$\hspace*{5cm}\frac{dx}{dt}=-\alpha x + m_{1} f( x_{\tau_{1}})
+m_{2} f( x_{\tau_{2}}),\hspace*{5.3cm}(1)$$
$$\hspace*{5cm}\frac{dy}{dt}=-\alpha y + m_{3} f( y_{\tau_{1}})
+ m_{4} f (y_{\tau_{2}}) + K f (x_{\tau_{3}}),\hspace*{3.5cm}(2)$$
where $f$ is differentiable generic nonlinear function. Throughout this 
paper $x_{\tau}\equiv x(t-\tau)$.\\ 
One finds that under the condition 
$$\hspace*{6cm}m_{1}-K=m_{3}, m_{2}=m_{4}\hspace*{6.5cm}(3)$$
Eqs. (1) and (2) admit the synchronization manifold 
$$\hspace*{7cm}y=x_{\tau_{3}-\tau_{1}}.\hspace*{7.9cm}(4)$$
This follows from the dynamics of the error $\Delta=x_{\tau_{3}-\tau_{1}}-y$ 
$$\hspace*{5cm}\frac{d\Delta}{dt}= -\alpha\Delta + m_{3} \Delta_{\tau_{1}} 
f'( x_{\tau_{3}}) + m_{2} \Delta_{\tau_{2}} f' (x_{\tau_{2}+\tau_{3}-\tau_{1}}).\hspace*{2.7cm}(5)$$
Here $f'$ stands for the derivative of $f$ with respect to time and the derivative should 
be bounded. The sufficient stability condition of the trivial solition $\Delta=0$ of (5) can be 
found using Razumikhin theorems (see for details,[5],pp.151-161). In [5] considering the 
Lyapunov function $V=\frac{1}{2}\Delta ^2$ and using Razumikhin theorems (hereafter we 
use the term Razumikhin-Lyapunov approach), 
under restrictive inital data condition 
$\vert \Delta(t) \vert \geq \vert \Delta(t+\theta)\vert$, for all $\theta \in [-\tau_{j} ,0]$ it 
is shown that zero solition of $\frac{d\Delta}{dt}=-a(t)\Delta(t) -\sum_{j=1}^{j=n} b_{j}(t)\Delta(t-\tau_{j}(t))$
is uniformly asymptotically stable for all bounded continous functions $a,b_{j},\tau_{j}$ 
if $a(t)\geq\delta >0,\sum_{j=1}^{j=n} b_{j}(t) <p\delta , 0<p<1, 0<\tau_{j}(t)<\tau $ 
for all $t\in (-\infty, \infty)$. Thus, applying the Razumikhin-Lyapunov 
approach we obtain that the sufficient stability 
condition for the synchronization manifold (4) can be written as:
$$\hspace*{5cm}\alpha > \vert m_{3}(\sup f'( x_{\tau_{3}}))\vert +\vert m_{2}(\sup f' (x_{\tau_{2}+\tau_{3}-\tau_{1}})) \vert.\hspace*{2.9cm}(6)$$ 
Here $\sup f'(x)$ stands for the supremum of $f'$ with respect to the appropriate norm.\\
We notice that 
for $\tau_{3}>\tau_{1}$,$\tau_{3}=\tau_{1}$, and $\tau_{3}<\tau_{1}$ 
 (4) is the retarded, complete, and anticipating synchronization manifold [7,8], 
respectively.Analogously one finds both the existence ($m_{2}-K=m_{4}$, $m_{1}=m_{3}$)
and sufficient stability 
($\alpha > \vert m_{3}(\sup f'( x_{\tau_{3}}))\vert +\vert m_{2}(\sup f' (x_{\tau_{2}+\tau_{3}-\tau_{1}}))\vert $) 
conditions for synchronization  manifold $y=x_{\tau_{3}-\tau_{2}}$.\\
In the case of linear coupling of form $K(x-y)$ one obtains that under the condition 
$m_{1}=m_{3}, m_{2}=m_{4}$ synchronization manifold $y=x$ exists and it is stable 
if $\alpha + K > \vert m_{1}(\sup f'( x_{\tau_{1}}))\vert +\vert m_{2}(\sup f' (x_{\tau_{2}})) \vert.$ 
For the parameter mismatches, e.g. $\tau_{1} \neq \tau_{1f}$, ($\tau_{1f}$ is the 
delay time for the first feedback loop in $y$ dynamics) it is 
clear that complete synchronization is not the synchronization manifold. 
Then for such a case we use the auxiliary system method to 
detect generalized synchronization [9]:that is given another identical driven auxiliary system $z(t)$, generalized synchronization between 
$x(t)$ and $y(t)$ is established with the achievement of complete synchronization 
between $y(t)$ and $z(t)$. Thus,the auxiliary method allows to find the local stability 
condition of the generalized synchronization [9]. Applying this method we find local 
stability condition of the generalized synchronization between $y$ and $x$:
$\alpha + K > \vert m_{3}(\sup f'( y_{\tau_{1f}}))\vert + \vert m_{4}(\sup f'( y_{\tau_{2}}))\vert.$ 
We also notice for $Kx_{\tau_{3}}$ type of 
coupling with mismatches between 
the relaxation coefficents the only possible synchronization manifold is $y=x_{\tau_{3}}$ 
(independent of the relation between the coupling and feedback delay times)
with existence $\alpha_{2}-\alpha_{1} =K$,$m_{1}=m_{3}$, and $m_{2}=m_{4}$ and sufficent 
stability conditions  
$\alpha_{2} > \vert m_{1}(\sup f'( x_{\tau_{1} + \tau_{3}}))\vert +\vert m_{2}(\sup f' (x_{\tau_{2}+\tau_{3}}))\vert .$ 
The presence of such a mismatch could be useful in the interpretation of the future 
experiments with coupling-delay lag synchronization and could serve as a "switch off" 
mechanizm for certain types of synchronization manifolds.\\
Generalization of the approach to $n$-tuple feedback systems, i.e. systems with 
multiple delays of type (1) and (2) is straightforward. We underline that a stability 
condition derived from the Razumikhin-Lyapunov approach is a sufficient condition: it 
assures a high quality synchronization for a coupling strength estimated from the 
stability condition, but does not forbid the possibility of synchronization with smaller 
coupling strengths. The threshold coupling strength can be estimated by the dependence of 
the maximal Lyapunov exponent $\lambda$ of the error dynamics on K:i.e. 
from $\lambda (K)=0$ [10].\\
\indent {\it Example 1:The Ikeda model.}- First we test the approach on the nonlinearly 
coupled Ikeda model:
$\frac{dx}{dt}=-\alpha x + m_{1} \sin x_{\tau_{1}}
+m_{2} \sin x_{\tau_{2}};\frac{dy}{dt}=-\alpha y + m_{3} \sin y_{\tau_{1}}
+ m_{4} \sin y_{\tau_{2}} + K \sin x_{\tau_{3}},$
with positive $\alpha_{1,2}$ and $-m_{1,2,3,4}$.This investigation is of 
considerable practical importance, as the equations of 
the class B lasers with feedback (typical representatives of class B are solid-state, 
semiconductor, and low pressure $CO_{2}$ lasers [11]) can be reduced to an equation of the 
Ikeda type [12].The Ikeda model was introduced to describe the dynamics of an optical bistable resonator, 
plays an important role in electronics and physiological studies and is well-known for 
delay-induced chaotic behavior [7,13]. We find that the Ikeda systems can be synchronized on the 
synchronization manifold $y=x_{\tau_{3}-\tau_{1}}$ under the condition $m_{1}-K=m_{3}, m_{2}=m_{4}$
and is stable if $\alpha > \vert m_{3} \vert +\vert m_{2} \vert.$ 
Numerical simulations fully support the analytical results. The Ikeda model 
was simulated using the DDE23 program [14] in MATLAB 6. 
Figure 1 shows the time series of the 
driver $x(t)$ (solid line) and the driven system $y(t)$(dotted line) for $\alpha =5$, 
$\tau_{1}=3$,$\tau_{2}=2$,$\tau_{3}=1$, $m_{2}=m_{4}=-1$,$m_{1}=-18$, $m_{3}=-1$ and 
$K=-17$. After transients the driven system shifted $\tau_{3}-\tau_{1}=-2$ time units to 
the left and $y=x(t+2)$ (anticipating synchronization).\\
\indent{\it Example 2: The Mackey-Glass model.}-Consider complete synchronization between 
the linearly coupled double-feedback Mackey-Glass systems:
$\frac{dx}{dt}=-\alpha_{1} x + 
k_{1} \frac{x_{\tau_{1}}}{1+x_{\tau_{1}}^{b}} +
k_{2} \frac{x_{\tau_{2}}}{1+x_{\tau_{2}}^{b}};\frac{dy}{dt}=-\alpha_{2} y +k_{3} 
\frac{y_{\tau_{1}}}{1+y_{\tau_{1}}^{b}} +k_{4} 
\frac{y_{\tau_{2}}}{1+y_{\tau_{2}}^{b}}
+ K(x-y).$The dynamical variable in the Mackey-Glass model is the concentration of the mature cells 
in blood at time $t$ and the delay time is the time between the initiation of cellular 
production in the bone marrow and the release of mature cells into the blood [1,10].(At 
present there is also an electronic analog of the Mackey-Glass system [10].)
We find that the Mackey-Glass systems can be synchronized on the synchronization 
manifold $y=x$ under the existence $k_{1}=k_{3}, k_{2}=k_{4}$ and stability 
$\alpha + K > (k_{1} +k_{2})\frac{(b-1)^{2}}{4b}$ conditions. For analytical estimations 
we took into account that the absolute maximum of the function $ \vert f^{\prime} (x_{\tau}) \vert $  is obtained
at $x_{\tau}=(\frac{b+1}{b-1})^{\frac{1}{b}}$ and is equal to $ 
\frac{(b-1)^{2}}{4b}$ [10]. 
Figure 2 shows numerical simulation of the linearly coupled Mackey-Glass models:time 
series of the driver $x(t)$ (solid line) and driven system $y(t)$ (dotted line).
The parameters are 
$\tau_{1}=14,\tau_{2}=20,\alpha =1,b=10,k_{1}=k_{3}=2,k_{2}=k_{4}=0.2,K=5.$ 
After transients the driven systems trajectory completely coincides with that of the 
driver system. 
In Figure 3 generalized synchronization between the linearly coupled Mackey-Glass 
models is shown for $\tau_{1}=14,\tau_{1f}=16$;the other parameters are as in Fig.2.\\
We emphasize that as the coupling strength estimated from the 
stability condition gives a high-quality synchronization, the synchronization manifold 
is robust against perturbations of the coupling strength. As mentioned above the onset 
of synchronization occurs at the coupling stength when the maximal Lyapunov exponent of 
the error dynamics vanishes as function of $K$ [10]. Our estimations show that for 
the parameters values as in Fig.2 the threshold value of K is:$K\approx -1.11$, which is 
far less than $K=3.44$ found from the sufficient stability condition.\\
\indent  {\it Example 3:The Lang-Kobayashi model.}-As the last example we study 
synchronization between the unidirectionally coupled double-feedback Lang-Kobayashi systems [1,6-8]
(laser diodes with a double external cavities, see [6] and references there-in):
$\frac{dE_{1,2}}{dt}= \frac{(1+\iota \alpha_{1,2})}{2} 
(G_{1,2}(N_{1,2}-N_{01,02}) -\gamma_{1,2})E_{1,2} + 
k_{1,2}E_{1,2}(t-\tau_{1})\exp(-\iota \omega \tau_{1}) + 
k_{3,4}E_{1,2}(t-\tau_{2})\exp (-\iota \omega \tau_{2})+
k_{5}E_{1}(t-\tau_{3})\exp (-\iota \omega \tau_{3});\frac{dN_{1,2}}{dt}=
J_{1,2} -\gamma_{e1,e2}N_{1,2}- 
G_{1,2}(N_{1,2}-N_{01,02})\vert E_{1,2} \vert ^{2},$
where $E_{1,2}$ are slowly varying complex fields for the master and slave lasers, 
respectively $N_{1,2}$ are the carrier densities;$N_{01,02}$ are the carrier 
densities at transparency;$\gamma_{1,2}$ are the cavity losses;
$\alpha_{1,2}$ are the linewidth enhancement factors; $G_{1,2}$ are the optical gains;
$k_{1,2}$ and $k_{3,4}$ are the feedback levels for the master and slave lasers, 
respectively.$k_{5}$ is the coupling rate;$\omega$ is the optical feedback frequency 
without feedback; $\tau_{1,2}$ are the round-trip times in the external cavities for 
the coupled lasers; $\tau_{3}$ is the time flight between the master laser and slave 
laser (coupling delay time);$J_{1,2}$ are the injection currents; $\gamma^{-1}_{e1,e2}$ 
are the carrier lifetimes.The term $k_{5}$ exists only for the slave laser. In order to 
find possible synchronization regimes we compare e.g. equations for the 
dynamics of $E_{2}$ and $N_{2}$ with dynamics  
of $E_{1,\tau_{3}-\tau_{1}}$ and $N_{1,\tau_{3}-\tau_{1}}$ and find that for the case of 
identical lasers (all parameters are the same except for the feedback and coupling rates)
 Lang-Kobayashi systems can be synchronized on the 
synchronization manifold $I_{2}=I_{1,\tau_{3} -\tau_{1}}$
if 
$$\hspace*{6cm}k_{1}=k_{3} + k_{5}, k_{2}=k_{4}\hspace*{7.1cm}(7)$$
as the intensities $I_{2}$ and $I_{1,\tau_{3}-\tau_{1}}$ ($I=\vert E \vert ^{2}$) 
can be made identical under these conditions. We have found the same existence conditions 
for the unidirectionally coupled laser diodes with incoherent optical  
feedbacks [15]. Unfortunately the stability of the synchronization manifolds for the 
Lang-Kobayashi model practically can not be studied analytically, 
and therefore one have to rely on the numerical methods. 
Figure 4 shows synchronization manifold $I_{2}$ vs.$I_{1,\tau_{3} -\tau_{1}}$ 
for parameters values $N_{0}=1.7\times 10^{8}$,$G=2.14\times 10^{4}$,
$\tau_{1}=10^{-8}s$,$\tau_{2}=1.5\times10^{-8}s$,$\tau_{3}=2\times 10^{-8}$,
$\alpha =5$,
$\gamma_{e} =0.9\times 10^{-9}s$, $\frac{2\pi}{\omega}=635nm$,$J=0.02\gamma N_{0}$, 
$k_{1}=10 ns^{-1}$,$k_{3}=1ns^{-1}$,$k_{5}=9 ns^{-1}$,$k_{2}=k_{4}=100 ns^{-1}$.\\
\indent {\it Conclusions.}- By using the Razumikhin-Lyapunov approach we have 
presented the first report on different synchronization regimes between two unidirectionally 
coupled (linearly and nonlinearly) chaotic systems with multiple delays. 
We have successfully applied the approach to the paradigm models in nonlinear physics-the 
Ikeda, Mackey-Glass and Lang-Kobayashi models. We have found analytically the existence 
and whenever possible stability conditions for the anticpating,lag, complete and 
generalized synchronization regimes. We hope that this research opens up possiblities for 
highly anticipated intensive experimental investigations of chaos synchronization in 
systems with multiple time-delays.\\
The author acknowledges support from the Abdus Salam ICTP (Trieste, Italy) Associate scheme.\\
\begin{center}
Figure captions
\end{center}
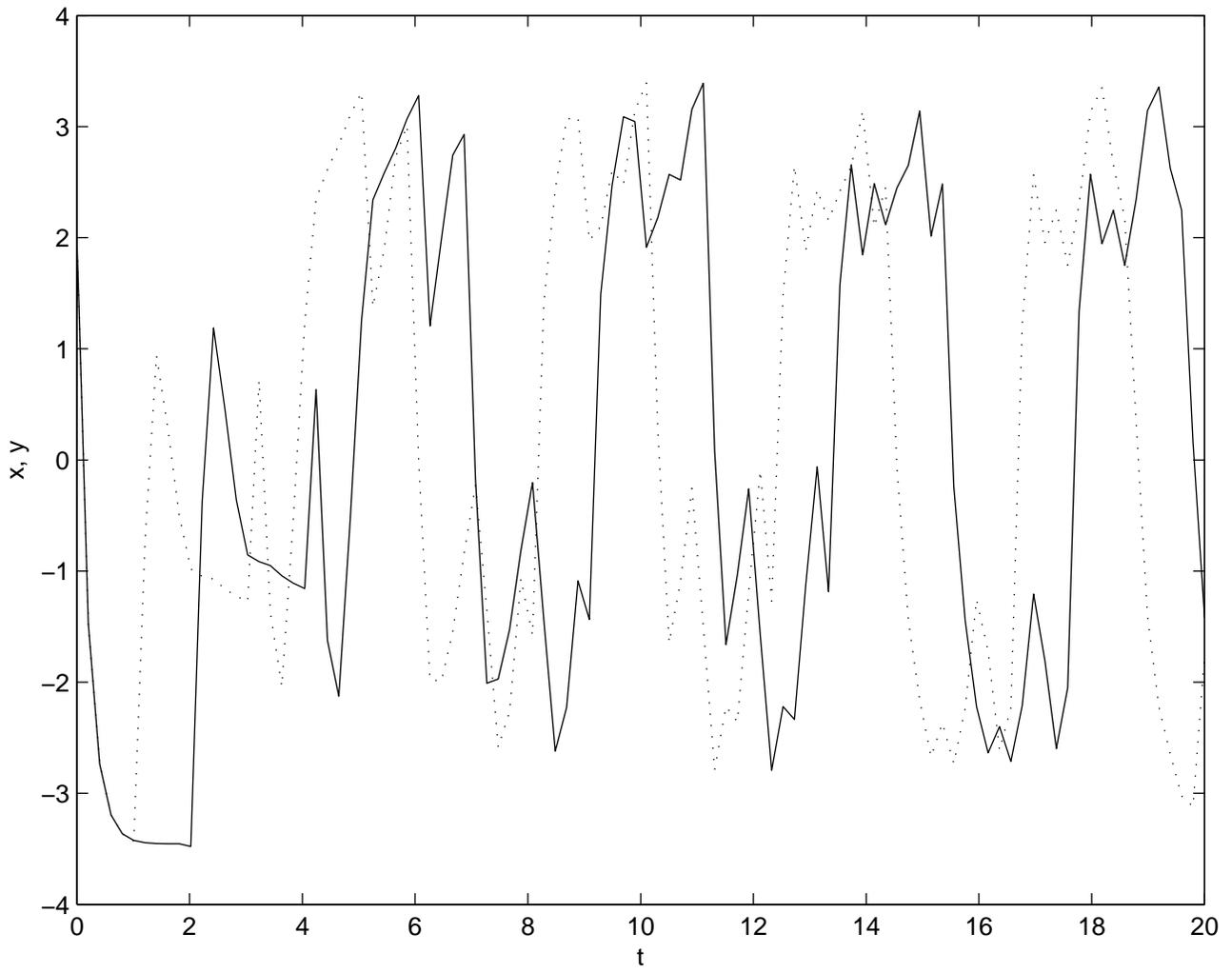
\noindent FIG.1.Numerical simulation of the Ikeda model:
the time series of the driver $x(t)$ (solid line) and the driven system 
$y(t)$(dotted line);$y=x(t +\tau_{3}-\tau_{1})=x(t+2).$ Dimensionless units.\\
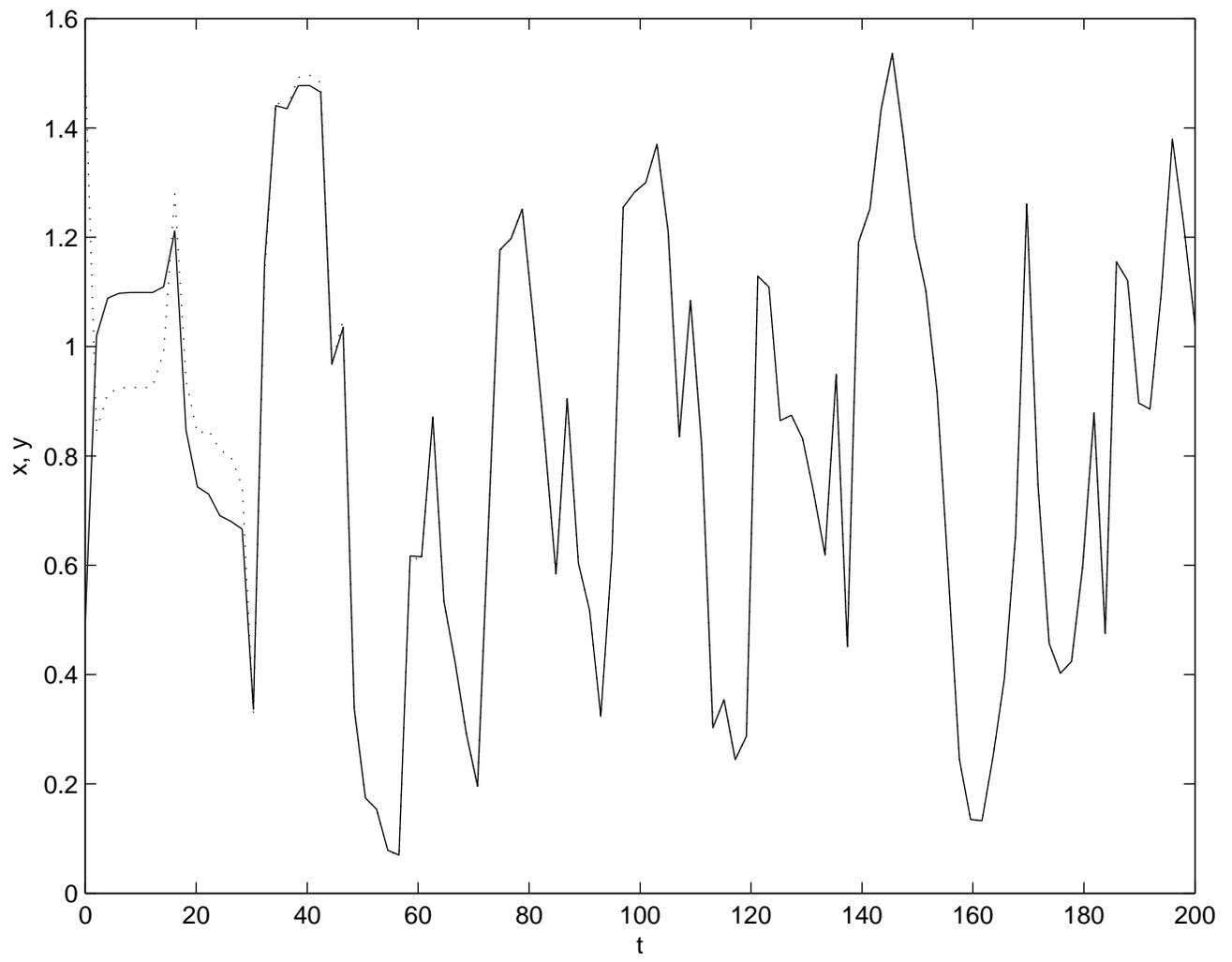
FIG.2.Numerical simulation of the Mackey-Glass model:
complete synchronization between $y$ and $x.$Dimensionless units.\\
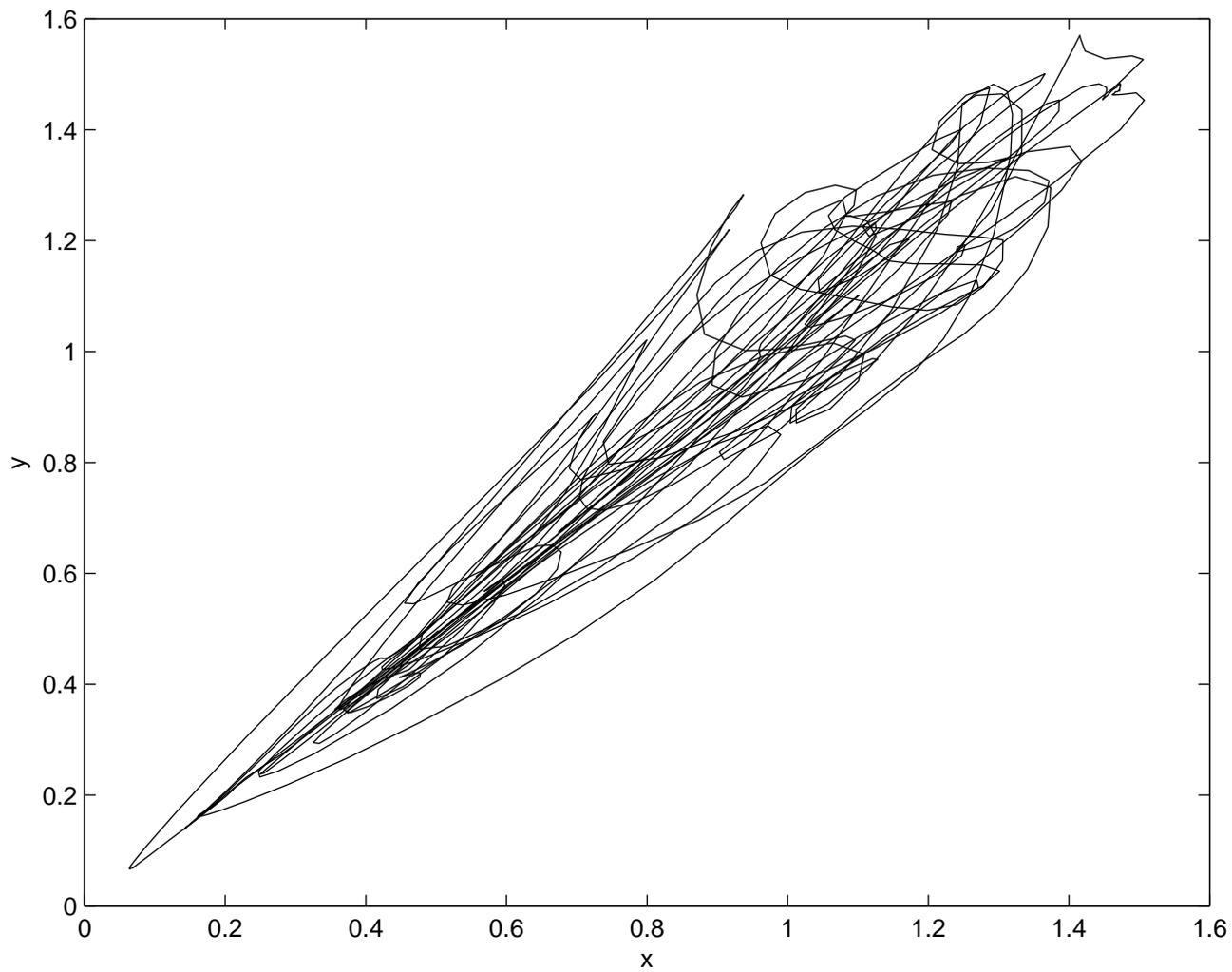
FIG.3.Numerical simulation  of the Mackey-Glass model: generalized synchronization 
between $y$ and $x$.Dimensionless units.\\
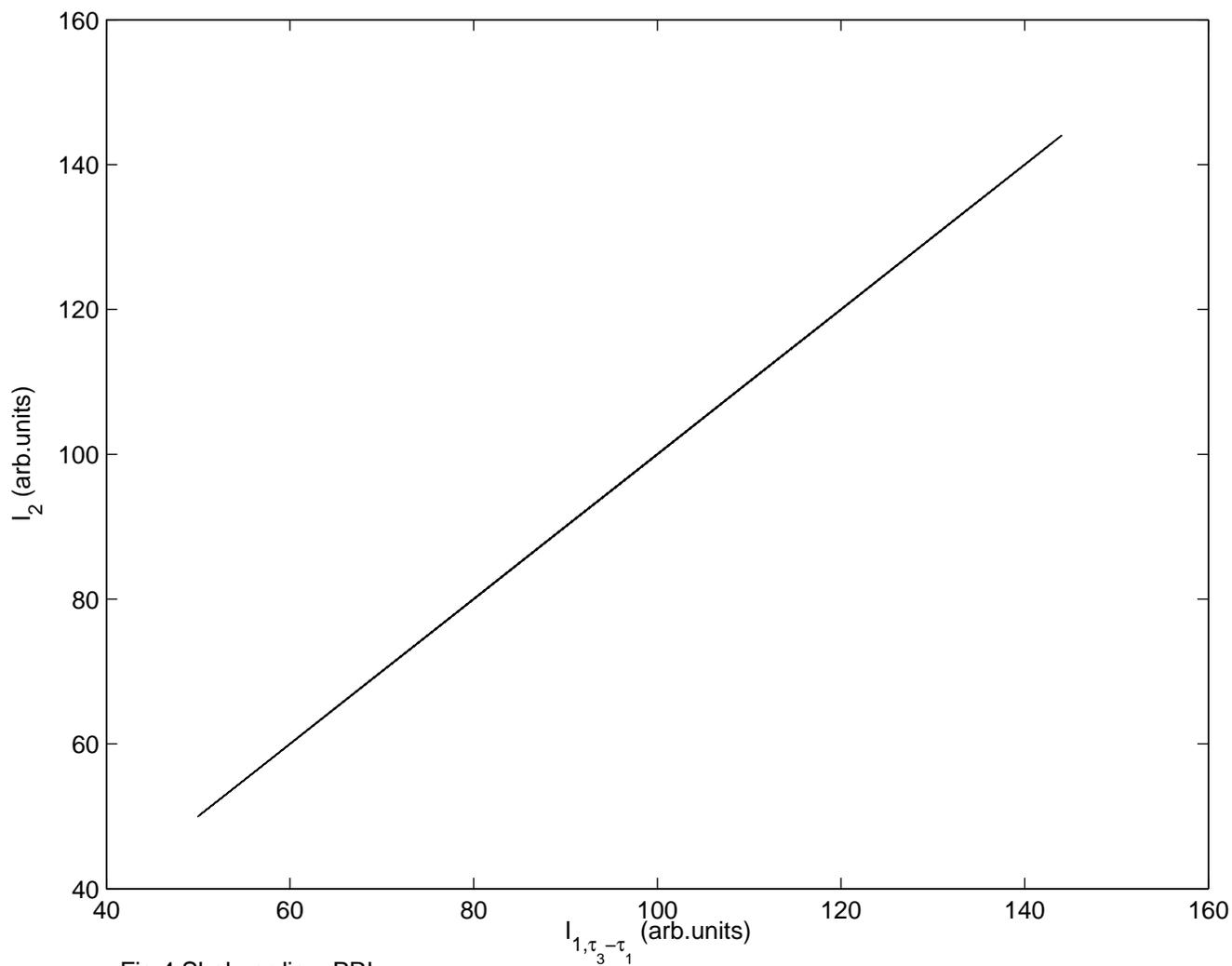
FIG.4. Numerical simulation of the Lang-Kobayashi model:the dependence of $I_{2}$ 
on $I_{1,\tau_{3}-\tau_{1}}$.\\


\begin{thebibliography}{99}
\bibitem{} L. M. Pecora and T. L. Carroll, Phys. Rev. Lett. {\bf 64}, 821 (1990);
E.Ott, C.Grebogi, and J.A.Yorke, Phys.Rev.Lett. {\bf 64}, 1196 (1990).\\
\bibitem{} S.J.Boccaletti {\it et al}, Phys.Rep.{\bf 366}, 1 (2002); Focus Issie:Control 
and synchronization in chaotic dynamical systems [Chaos {\bf 13},126 (2003)].\\
\bibitem{} R.Dilao,T.Domingos, and E.M.Shahverdiev,Mathematical Biosciences {\bf 189},141(2004).\\
\bibitem{} K.Ikeda,H.Daido, and O.Akimoto,Phys.Rev.Lett.{\bf 45},709 (1980);M.C.Mackey and L.Glass,Science {\bf 197},287 (1977);
R.Lang and K.Kobayashi,IEEE J.Quantum.Electron.{\bf QE-16},347(1980);E.Tziperman {\it al},
Science {\bf 264},72 (1994);P.K.Asea and P.J.Zak,J.Econ.Dyn.Control B, {\bf 23},1155 (1999);
G.D.VanWiggeren and R.Roy,Science {\bf 279},1198 (1998);F.M.Atay,J.Jost, and A.Wende,
Phys.Rev.Lett.{\bf 92},(2004).\\
\bibitem{} J.K.Hale and S.M.V.Lunel, Introduction to Functional Differential 
Equations (Springer, New York, 1993).\\
\bibitem{} Y.Liu and J.Ohtsubo, IEEE J.Quantum Electron. {\bf 33}, 1163 (1997);
 F.Rogister, P.Megret and M.Blondel, Phys.Rev. E {\bf 67}, 027203 (2003).\\
\bibitem{} C.Masoller, Phys.Rev.Lett. {\bf 86},2782(2001);
 C.Masoller and D.H.Zanette, Physica A {\bf 300},359 (2001).\\
\bibitem{} E.M.Shahverdiev, S.Sivaprakasam, and K.A.Shore, Phys.Rev.E  {\bf 66}, 017204 (2002).\\
\bibitem{} M.Zhang {\it et al}, Phys.Rev.E {\bf 68},036208 (2003).\\
\bibitem{} K.Pyragas,Phys.Rev.E {\bf 58}, 3067 (1998).\\
\bibitem{} Ya.I.Khanin, Chaos {\bf 6}, 373 (1996).\\
\bibitem{} F.T.Arecchi, G.Giacomelli, A.Lapucci, and R.Meucci, Phys.Rev.A {\bf 43},4997 (1994).
\bibitem{} H.U.Voss, Phys.Rev.E {\bf 61}, 5115 (2000).\\
\bibitem{} L.F.Shampine and S.Thompson, Appl.Numer.Math.{\bf 37},441 (2001).\\
\bibitem{} F.Rogister {\it et al} Optics Letters {\bf 26},1486 (2001).\\
\end{thebibliography}
\end{document}